\documentclass{osa-article}

\journal{osajournal}

\articletype{Research Article}
\usepackage{siunitx}
\usepackage{float}
\usepackage{nicefrac}
\usepackage{cite}
\usepackage[T1]{fontenc}
\usepackage[utf8]{inputenc}
\usepackage{titlesec}
\begin{document}

\title{Characterizing ultrashort laser pulses with second harmonic dispersion scans}

\author{Ivan Sytcevich,\authormark{1,6}\authormark{*}, Chen Guo\authormark{1,6}, Sara Mikaelsson\authormark{1}, Jan Vogelsang\authormark{1,2}, Anne-Lise Viotti \authormark{1}, Benjamín Alonso\authormark{3}, Rosa Romero\authormark{4}, Paulo T. Guerreiro\authormark{4}, Anne L'Huillier\authormark{1}, Helder Crespo\authormark{4,5},  Miguel Miranda\authormark{1,4}, and Cord L. Arnold\authormark{1}}

\address{\authormark{1}Department of Physics, Lund University, Box 118, 22100 Lund, Sweden\\
\authormark{2}NanoLund, Lund University, Box 118, 22100 Lund, Sweden\\
\authormark{3}Grupo de Aplicaciones del Láser y Fotónica (ALF), Departamento de Física Aplicada, University of Salamanca, Plaza de la Merced s/n 37008 Salamanca, Spain\\
\authormark{4}Sphere Ultrafast Photonics, Rua do Campo Alegre, 1021, Edifício FC6, 4169-007 Porto, Portugal\\
\authormark{5}IFIMUP-IN and Departamento de Física e Astronomia, Universidade do Porto, Rua do Campo Alegre 687, 4169007 Porto, Portugal \\
\authormark{6}These authors contributed equally to this work }

\email{\authormark{*}ivan.sytcevich@fysik.lth.se} 

\titleformat*{\section}{\large\bfseries}
\titleformat*{\subsection}{\normalsize\itshape}


\begin{abstract}
The dispersion scan (d-scan) technique has emerged as a simple-to-implement characterization method for ultrashort laser pulses. D-scan traces are intuitive to interpret and retrieval algorithms that are both fast and robust have been developed to obtain the spectral phase and the temporal pulse profile.
Here, we give a review of the d-scan technique based on second harmonic generation. We describe and compare recent implementations for the characterization of few- and multi-cycle pulses as well as two different approaches for recording d-scan traces in single-shot, thus showing the versatility of the technique. 
\end{abstract}

\section{Introduction}

Ultrashort laser pulses have become an indispensable tool in numerous fields of science and engineering and found multiple applications in physics, chemistry, material processing and medicine. 
Almost directly after the invention of the laser, the introduction of passive mode-locking techniques led to light pulses with durations in the picosecond range \cite{LambPR1964,IppenAPL1972}. 
The discovery of Titanium:Sapphire as a laser-active material in the middle of the eighties \cite{MoultonJOSAB1986}, together with chirped pulse amplification (CPA) \cite{StricklandOC1985} and Kerr-lens mode-locking \cite{SpenceOL1991} resulted in rapid commercialisation and spread of the technology. Advanced nonlinear post-compression techniques \cite{ForkOL1987,NisoliAPL1996,NisoliOL1997} opened up for pulses with durations down to only few femtoseconds in the visible and near infrared spectral regions. In this regime, the pulse envelope contains only a few oscillations of the electric field which gives access to a variety of exciting physical phenomena \cite{PaulusNature2001,Schiffrin2013}. Such ultrashort pulses can be used to produce even shorter waveforms by the process of High-order Harmonic Generation \cite{McPhersonJOSAB1987,FerrayJPB1988}, which further pushes achievable pulse widths down to the attosecond regime \cite{PaulScience2001,HentschelN2001,GaumnitzOE2017}, allowing experimental studies with unprecedented time resolution. 

Many applications of ultrashort laser pulses require their accurate characterization, i.e. the determination of the exact waveform of the laser pulse or at the least its pulse intensity profile. Both are challenging tasks, since it is not easy to directly access the pulse information in the time domain.
Direct time-resolved diagnostics, e.g. streaking measurements \cite{GoulielmakisS2004} and electro-optic sampling-based approaches \cite{KeiberNP2016} have been demonstrated. These techniques, however, require powerful laser pulses and complex setups. Less demanding experimental approaches have been proposed to characterize ultrashort pulses. The intensity autocorrelation measurement was one of the first techniques to be introduced \cite{Ippen1977} and is still widely used. It records the intensity of a nonlinear signal (usually second harmonic) as a function of the delay between two pulse replicas to obtain an estimate of the duration of the pulse temporal profile. The exact pulse amplitude and the phase information remain however unavailable \cite{ChungIJostiqe2001}.
By adding a spectrometer to the detection scheme and measuring a spectrum at each delay, a two-dimensional spectrogram can be obtained which is the basis of the Frequency Resolved Optical Gating (FROG)\cite{KaneJQE1993,TrebinoJOSAA1993} technique. With the use of iterative mathematical algorithms both phase and amplitude can be retrieved and the pulse fully reconstructed. Another popular approach, named Spectral Phase Interferometry for Direct Electric-field Reconstruction (SPIDER) \cite{IaconisOL1998,WyattOL2006}, relies on recording a spectral interference pattern between two delayed and frequency-sheared pulse replicas. Compared to FROG, this method does not require complex retrieval algorithms, at the expense of a more complicated optical setup.

A different class of characterization techniques does not rely on pulse replicas, but manipulates the pulse in the spectral domain.
In Multiphoton Intrapulse Interference Phase Scan (MIIPS), a spectral phase shaper is used to apply controlled phase functions to the pulse while the second harmonic spectrum is measured \cite{LozovoyOL2004}. The Group Delay Dispersion (GDD) curve can be obtained by establishing which function locally cancels out the original spectral phase and therefore maximizes the second harmonic generation (SHG) output at each wavelength, thus allowing the retrieval of the spectral phase and consequently the reconstruction of the temporal pulse profile. Besides MIIPS, related approaches utilizing pulse shapers have been reported \cite{GrabielleCaE2CD2009,LoriotOE2013}.

The dispersion scan, d-scan in short, utilizes a concept that is strongly related to MIIPS \cite{MirandaOE2012,MirandaOE2012a}.
A spectral phase is applied to the pulse to be characterized, by introducing a variable dispersive element, e.g. a glass wedge pair or a prism/grating compressor. By changing the amount of dispersion, e.g. by moving glass wedges of variable thickness in and out of the beam and recording the spectrum of a nonlinear signal (second harmonic, for instance), a two-dimensional trace is produced from which the phase information can be obtained with iterative algorithms following similar strategies as for FROG retrieval.
The immediate advantage of the d-scan technique is the simple setup without the need for pulse replicas or spectral shearing.
Furthermore, d-scan often uses a compressor to manipulate the spectral phase, an essential building block of almost any ultrafast laser, and thus allows for simultaneous compression and characterization of ultrashort light pulses.
Since its invention, the d-scan has become a well-established technique in many laboratories around the world. It has been implemented and tested with different target pulse widths and central frequencies and d-scan-compressed pulses have enabled a variety of applications ranging from pump-probe spectroscopy to biomedical imaging \cite{GoncalvesSR2016,MaibohmBOE2019}.

In this paper, we give a brief tutorial of the main features of the d-scan technique and we present an overview of recent developments and results obtained at the Lund Laser Centre (LLC).
In section \ref{sec:theory}, we provide a basic theoretical description and introduce the mathematical framework needed to describe a d-scan measurement. We give insights on how to interpret d-scan traces and how to choose phase retrieval strategies. In Section \ref{sec:scanning}, we describe different experimental implementations of the technique, for a wide range of pulse durations and wavelengths. Next, we present single-shot methods and discuss the advantages and limitations of using d-scan as a single-shot technique (section \ref{sec:siscan}). Finally, we conclude and give an outlook towards future developments of the method.


\section{Theory}
\label{sec:theory}

\subsection{The concept of a d-scan measurement}

We first provide a simple theoretical description and discuss generic properties of d-scan measurements. This will further help with the understanding of the advantages and limitations of this characterization technique as well as the reasoning behind certain engineering solutions.

The complex electric field representing a laser pulse can be expressed in the frequency domain as:
\begin{equation}
    \tilde{U}(\omega) = |\tilde{U}(\omega)| \exp{[i\phi(\omega)]} = \int_{-\infty}^{\infty}U(t) \exp{(-i\omega t)}dt,
    \label{eq:pulsefreq}
\end{equation}
where $|\tilde{U}(\omega)|$ is the spectral amplitude, $\phi(\omega)$ is the spectral phase and $U(t)$ is the corresponding complex electric field in the time domain. Propagating the pulse through a transparent medium of thickness $z$ is equivalent to multiplying Eq.(\ref{eq:pulsefreq}) with a phase term:

\begin{equation}
    \tilde{U}(z,\omega) = |\tilde{U}(\omega)| \exp{[i\phi(\omega)]} \cdot \exp{[ik_0(\omega)n(\omega)z]},
\end{equation}
where $n$ is the refractive index of the medium and $k_0$ is the vacuum wavenumber.

Pulse measurement techniques usually employ a nonlinear process in order to obtain pulse amplitude and phase sensitivity. Mathematically, the result of a nonlinear interaction can be written as
\begin{equation}
    U_\mathrm{NL}(z,t) = f\left[\int_{-\infty}^{\infty}\tilde{U}(z,\omega) \exp{(i\omega t)}d\omega\right],
\end{equation}
where $f$ represents the particular nonlinear interaction. In this article, we mainly deal with second harmonic generation (SHG) d-scan, where $f$ simply stands for squaring. Finally, the power spectrum of the process is measured as a function of dispersion and a 2D trace is obtained:
  \begin{equation}
     I(z,\omega) = \left\lvert\int_{-\infty}^{\infty}U_\mathrm{NL}(z,t) \exp{(-i\omega t)}dt\right\rvert^2.
 \end{equation}
 
The simple model presented above assumes an ideal coupling of the fundamental radiation to the nonlinear signal, which implies perfect phase matching over the pulse bandwidth. For broadband few-cycle pulses, it is usually not the case \cite{WeinerIJoQE1983,BaltuskaIJoQE1999} and a response function $R(\omega)$ (which may contain not only the effect of finite phase matching, but also technical parameters, e.g. a spectrometer response function) has to be included to accommodate for the irregular spectral response,
\begin{equation}
    I_\mathrm{real}(z,\omega) = R(\omega)\cdot I_\mathrm{ideal}(z,\omega).
    \label{eq:response}
\end{equation} 

\begin{figure}[hbt!]
     \centering
     \includegraphics[scale=0.8]{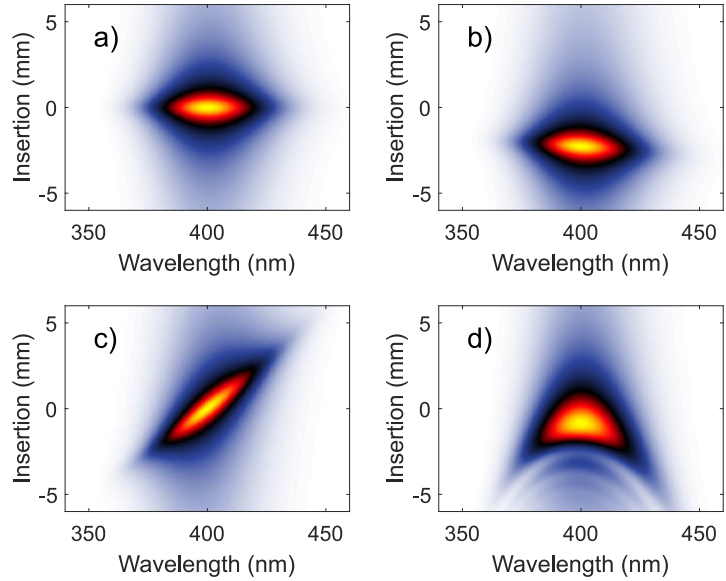}
     \caption{Simulated SHG d-scan traces for a 10 fs Gaussian pulse centered at 800 nm with no phase applied (a), 100 $\textrm{fs}^2$ GDD (b), 800 $\textrm{fs}^3$ TOD (c), 8000 $\textrm{fs}^4$ FOD (d).}
     \label{fig:traceinterp}
 \end{figure}

A second harmonic d-scan trace for an ideal \SI{10}{fs}-FWHM (Full Width at Half Maximum) Gaussian pulse with a center wavelength of \SI{800}{nm} is presented in Fig.~\ref{fig:traceinterp}(a). In this simulation, the index of refraction is calculated from the Sellmeier equation for BK7 glass, which is a common material used for d-scans wedges in the visible and near-infrared spectral ranges. In Fig.~\ref{fig:traceinterp}(b-d), we add numerically different dispersion orders in the Taylor expansion of the spectral phase, i.e. group delay dispersion (GDD), third-order phase (TOD) and fourth-order phase (FOD). 
Applying a positive GDD to the pulse mainly shifts the trace down along the dispersion axis (Fig.~\ref{fig:traceinterp}(b)), implying that the pulse can be re-compressed by removing the glass.
The trace appears to be slightly tilted, due to the fact that BK7 introduces not only GDD, but also higher-order terms. 
This becomes obvious in Fig.~\ref{fig:traceinterp}(c), where a d-scan trace with TOD, resulting in an almost linear tilt of the trace with respect to the dispersion axis, is shown.
Finally, FOD leads to a parabolic-like deformation (Fig.~\ref{fig:traceinterp}(d)).
These simple examples highlight the sensitivity of d-scan measurements to the spectral phase of the pulse. The d-scan trace therefore provides an intuitive way to visually estimate the quality of compressed pulses, even without using reconstruction algorithms, which is a very useful day-to-day optimization metric of few-cycle pulses from e.g. hollow-core fiber (HCF) compressors\cite{ConejeroJarqueSREP2018,SilvaOL2018}.

 \subsection{Phase retrieval} 

It is a straightforward procedure to calculate a d-scan trace for a known pulse. However, the reverse, i.e. extracting information \textit{from} a measured d-scan trace, is not such a trivial task. 
Mathematically, this falls into the class of inverse problems and is tackled by mathematical routines named phase retrieval algorithms.
The main idea is to find the pulse that generates a nearly-identical trace compared to the experimental data. Numerically, we seek to minimize a root mean square (RMS) error $G$ between the experimentally measured and the computed trace, sampled with $m = 1,2,...N_m$ points in frequency and $k = 1,2,...N_k$ different glass insertions:
\begin{equation}
    G^2 = \frac{1}{N_m N_k}\sum\limits_{m,k}\left(I_\mathrm{meas}(\omega_m,z_k)-\mu_m I_\mathrm{retr}(\omega_m,z_k)\right)^2.
    \label{eq:errorG}
\end{equation}
Here, $I_\mathrm{meas}$ and $I_\mathrm{retr}$ are the measured and simulated traces, respectively, and 
\begin{equation}
    \mu_m = {\sum\limits_{k}\left[I_\mathrm{meas}(\omega_m,z_k)I_\mathrm{retr}(\omega_m,z_k)\right]}\big/{\sum\limits_{k}I_\mathrm{retr}(\omega_m,z_k)^2},  
\end{equation}
is a minimization factor which is calculated and updated in every iteration. For a successful retrieval, $\mu_m$ gives the spectral response function $R(\omega)$ [Eq.~(\ref{eq:response})].
Equation \ref{eq:errorG} shows that pulse retrieval by minimizing $G$ essentially is a nonlinear least squares problem.
Solving such problems is a well-studied field of mathematics. Least-squares solvers like Nelder-Mead (NM), Levenberg-Marquardt (LM) or Broyden-Fletcher-Goldfarb-Shanno algorithms can be readily implemented as pulse retrievers and are used extensively with d-scan. The NM or downhill simplex, a method that was predominantly used in early d-scan works \cite{MirandaOE2012a,MirandaOE2012}, proved to be robust and reliable, albeit slow. The usage of LM-based minimization was reported in a self-calibrating d-scan technique, where the compressor parameters, i.e. introduced dispersion, could also be retrieved from the measurement \cite{AlonsoSREP2018}. This, in turn, allowed the quantification and elimination of pulse train instabilities in supercontinuum fiber lasers \cite{AlonsoSREP2020}. Another example is the d-scan retrieval algorithm based on differential evolution \cite{EscotoJOSAB2018}, which besides a faster convergence compared to NM was shown to be less prone to stagnate in local minima.
In general, in order to use this type of algorithms efficiently, it is beneficial to choose a convenient parametrization of the spectral phase. Expansion into a Fourier series usually increases the convergence speed, but in certain cases also the risk to get stuck in local minima. A possible workaround to this issue is to use a spline interpolation instead \cite{Baltuska1999} or to switch to a different basis whenever stagnation happens \cite{MirandaOE2012}. For long, "clean" pulses, which often have simple d-scan traces, a Taylor series representation of the phase can also be used.

Another class of retrieval algorithms, often predominantly used with FROG, is that of iterative constraint-based inversion algorithms (e.g. generalized projections or ptychography based approaches), inspired by early work in diffractive imaging \cite{GerchbergO1972}. The main feature of such methods is to introduce a set of certain constraints on the retrieved pulse in such a way that the error $G$ (eq. \ref{eq:errorG}) is reduced in each iteration. This arguably more elegant way of solving phase retrieval problems, is often faster compared to the "brute-force" minimization mentioned previously \cite{KaneIJSTQE1998,KaneJOSAB2008}. However, the speed-up often comes with the price of reduced robustness, especially when dealing with traces contaminated by noise. This was recently attributed to the fact that these algorithms do not converge to the least squares solution in the presence of Gaussian noise \cite{GeibO2019}. Thus, it might be preferable to choose general least squares solvers which were shown to be more reliable in these conditions \cite{WilcoxJOSAB2014}. To give an example, an algorithm based on data (or intensity) constraint was recently proposed for d-scan phase retrieval~\cite{MirandaJOSAB2017}. There, the data constraint means that the amplitude for the simulated complex d-scan trace is replaced with the measured data while the phase information is kept at each iteration of the algorithm. This method exhibits a faster convergence speed compared to the NM approach, but at the same time is significantly more susceptible to noise \cite{MirandaJOSAB2017}.

Generally speaking, the task of designing fast, robust and efficient retrieval algorithms is an active field of research and a significant amount of effort is devoted to the development of routines that are optimized for pulse characterization problems. The recently proposed COmmon Pulse Retrieval Algorithm (COPRA) \cite{GeibO2019}, for example, is a general algorithm that not only works with d-scan, but several other methods as well, like FROG or MIIPS. While being inspired by constraint-based methods, COPRA elegantly avoids the aforementioned problem of not reaching the least-squares solution by replacing a data constraint step with a gradient descent in the final stages of the algorithm run. This, in turn, helps to increase the accuracy of the retrievals for traces with high levels of Gaussian noise. Another exciting development is the use of artificial neural networks for pulse reconstruction \cite{ZahavyO2018}, which was recently reported for d-scan as well \cite{KleinertOL2019}, showing impressive ms-scale retrieval times and thus opening possibilities for a "Live View" pulse monitoring when combined with a single-shot d-scan system.


\section{Implementation}
 \label{sec:scanning}

After the first demonstration of the d-scan technique in 2012 \cite{MirandaOE2012}, which at that time focused on the characterization of pulses from few-cycle light sources in the near-infrared, there has been great effort to extend the applicability to pulses of different durations and central frequencies, combining various nonlinear phenomena with different approaches to introduce the required dispersion variation.

The most popular choice for the nonlinear interaction in d-scan measurements is SHG, owing to the availability of nonlinear media and high signal-to-noise ratio (SNR), e.g. when compared to third-order processes. However, the use of SHG can limit the applicability of d-scan in some situations. The limited phase-matching bandwidth of common SHG crystals usually reduces the effective spectral range of a single d-scan setup. The use of dielectric nanoparticles, which are free from phase-matching limitations, as a nonlinear medium was recently reported \cite{Perez-BenitoOL2019} as one possible solution. Another issue occurs when measuring pulses with octave-spanning spectra where there is an overlap between certain frequency components of the fundamental and second harmonic fields. In this case, the useful signal has to be carefully filtered e.g by using spatial masks or polarizers \cite{SilvaOE2014}. It is, however, worth mentioning that this usually undesirable feature can be beneficial: the produced interference of the fundamental and the SHG field is sensitive to the carrier-to-envelope phase and including this information in the retrieval algorithm allows for complete reconstruction of the electric field waveform \cite{MirandaOL2019}.

Higher-order nonlinear processes, e.g., Third Harmonic Generation (THG), can be utilized to alleviate these problems. D-scan setups based on THG in graphene \cite{Silva2013} and thin films of $\textrm{TiO}_2-\textrm{SiO}_2$ compounds \cite{HoffmannOE2014} have been reported. These materials have large nonlinear coefficients so that the problem of lower efficiency of third-order interactions is reduced. 
For pulses with spectral content extending towards the Ultraviolet (UV), approaches based on frequency up-conversion quickly become unpractical because of the need for specialized deep UV spectrometers. Another issue is the lack of suitable nonlinear crystals for efficient frequency conversion in the UV, as strong dispersion in this region prevents broad phase matching and as, in most materials, absorption becomes significant. To tackle this issue, schemes based on degenerate nonlinear processes have been introduced, where the frequency of the nonlinear signal is the same as the driving field. One of these schemes is Cross-Polarized Wave generation (XPW), which was successfully applied to d-scan for characterizing pulses in near-infrared \cite{TajalliOL2016} and deep UV \cite{TajalliOL2019}. Using XPW, it is crucial to have a high degree of linear polarization in the driving field and a polarization scheme with a large extinction ratio after XPW for detecting the signal with good SNR. Another degenerate process that was used in d-scan measurements is self diffraction\cite{CanhotaOL2017}, which enabled the simultaneous measurement of two unknown near-UV pulses\cite{CanhotaOL2017}.

\begin{figure}[hbt!]
    \centering
    \includegraphics{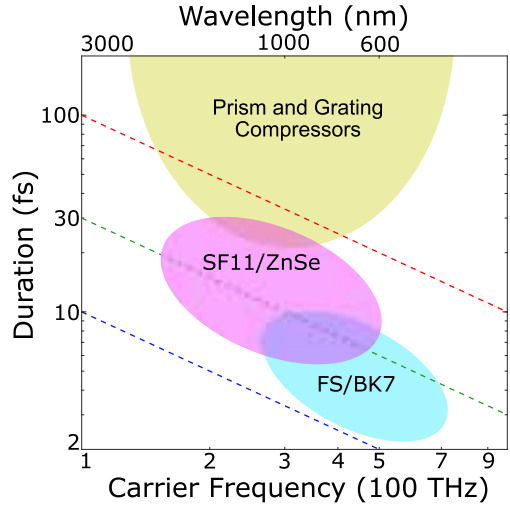}
    \caption{SHG d-scan implementations as a function of target pulse duration and central frequency. Blue, green and red dashed lines correspond to the duration of 1, 3 and 10 optical cycles at a given frequency, respectively. Shaded regions represent different optical components that can serve as a scanning dispersive element in a d-scan measurement. The light blue and pink areas correspond to a glass wedge pair configuration using the indicated materials.}
    \label{fig:vision}
\end{figure}

Even for the case of one selected nonlinear interaction, the experimental realization of a d-scan can still differ substantially depending on the central frequency and pulse duration (spectral bandwidth), as illustrated in Fig.~\ref{fig:vision} for SHG.
Generally, the longer the transform-limited pulse duration of the light source (the smaller the spectral bandwidth), the larger the dispersion scan window should be in order to capture the evolution of second harmonic around the point of optimal compression. For very short pulses, even small amounts of applied GDD result in significant compression/broadening, while for long pulses reaching ps widths or for pulses with large time-bandwidth product, the required GDD windows can be up to hundreds of thousands \si{fs^2}. How much dispersion, in terms of GDD window, exactly should be scanned in order to obtain robust measurement and retrieval is not a trivial question and requires a rigorous mathematical study which is outside of the scope of this paper. Here, we instead aim at giving practical values based on our experience when measuring pulses from different laser systems.

\begin{figure}[hbt!]
    \centering
    \includegraphics[scale=0.5]{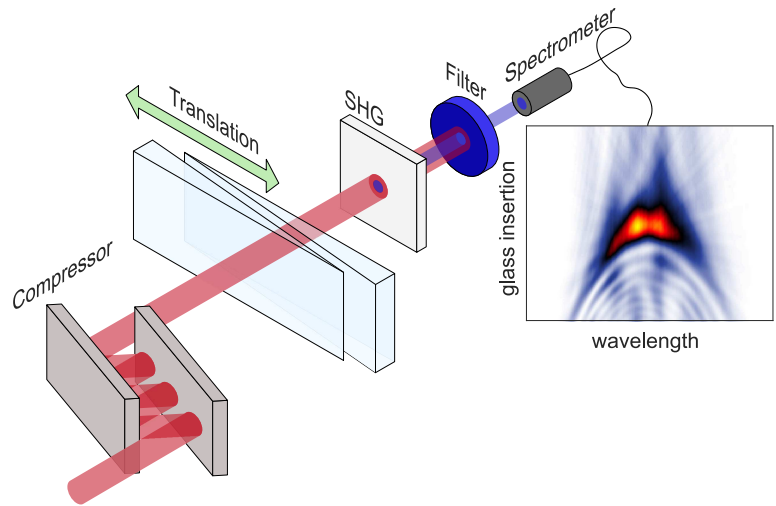}
    \caption{SHG-based d-scan setup for characterization of few-cycle pulses. The light passes through a compressor comprised of chirped mirrors and glass wedge pairs; The introduced GDD is finely tuned by the movement of one of the wedges. The second harmonic signal generated in the thin crystal is detected with a spectrometer and the trace is obtained by recording spectra at different wedge positions.}
    \label{fig:fewcyclesetup}
\end{figure}

The early designs, using wedges made of fused silica or BK7 glass with GDD in the range of 30-50~\si{fs^2/m\meter}, are well-suited for measuring few-cycle pulses with central frequencies in the visible and near-infrared (NIR), as emitted by hollow-core fiber (HCF) compressors or optical parametric chirped pulse amplification (OPCPA)-based lasers \cite{MirandaOE2012,BohleLPL2014,RudawskiEPJD2015} (Light blue shaded area in Fig. \ref{fig:vision}). A typical second harmonic d-scan setup is presented in Fig.~\ref{fig:fewcyclesetup}. After passing a chirped mirror pair, the pulse is usually negatively chirped. By finely tuning the insertion of the glass wedges and thus introducing positive GDD, the chirp can be controlled and contributions from optical components further down the beam path towards the experiment are compensated. A thin SHG crystal, a filter (to reject fundamental radiation) and a spectrometer are the only additional components needed to perform the measurement, making this configuration straightforward to implement. Additionally, since there is no beam splitting and recombining at any point, the required pulse energy to record a trace with a good SNR is very low, allowing measurements of the pulse directly from an oscillator. In the case of amplified pulses, the measurement can be done parasitically by using only a small portion of the energy of the main pulse (e.g. reflection off a glass plate/wedge). 

\begin{figure}[hbt!]
    \centering
    \includegraphics[width=\linewidth]{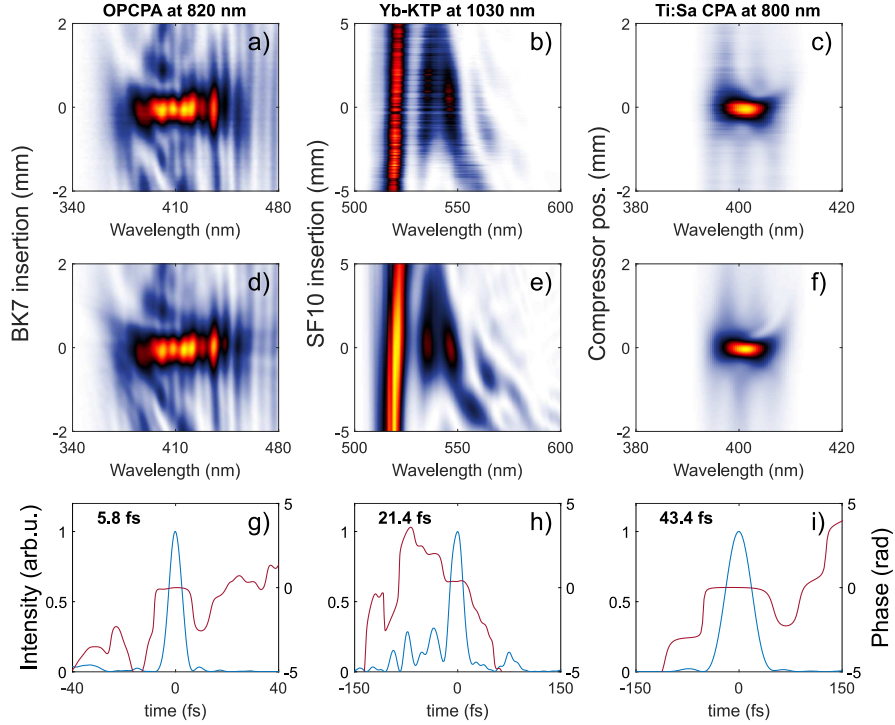}
    \caption{SHG d-scans in different pulse duration regimes: (a)-(c): measured traces using pulses from a few-cycle OPCPA system, from an Yb laser after post-compression in a KTP crystal \cite{ViottiCLEO2019} and from a 10 Hz CPA laser system, respectively; (d)-(f) corresponding retrieved traces; (g)-(i) retrieved pulse intensity profiles and phases.}
    \label{fig:difscan}
\end{figure}

A d-scan trace recorded from the output of a few-cycle high-repetition rate, Ti:Sapphire seeded, OPCPA laser \cite{HarthJO2017} located at LLC is shown in Fig.~\ref{fig:difscan}(a). A pair of BK7 glass wedges (about \SI{45}{fs^2/mm} of group velocity dispersion (GVD) at 800 nm) is used as a dispersive element, and a dispersion window of only \SI{180}{fs^2} is sufficient for the scan. The second harmonic is generated in a thin Beta Barium Borate (BBO) crystal. The fundamental radiation is filtered with a polarizer and the signal is recorded with a fiber-coupled spectrometer. Extracting the pulse information from the retrieved trace (Fig.~\ref{fig:difscan}(d)) gives a FWHM duration of \SI{5.8}{fs} (Fig.~\ref{fig:difscan}(g)).

When dealing with pulses having a central wavelength further to the infrared, it is often challenging and unpractical to introduce sufficient dispersion variation using wedges made from common optical glasses. With denser materials, e.g. SF10-SF57 flints, ZnS, ZnSe, etc. that have larger overall dispersion and zero-dispersion crossings further to the infrared (compared to standard glasses), the operating range of a standard d-scan setup can be extended to longer pulses (ca. \SI{20}{fs}) and wavelength regimes (<\SI{1.5}{\micro\meter}, indicated in pink in Fig.~\ref{fig:vision}). 

Fig.~\ref{fig:difscan}(b) shows a measured d-scan trace for pulses from a solid-state Ytterbium laser (\SI{1030}{nm} central wavelength) with a nonlinear post-compression stage in a single-domain potassium titanyl phosphate crystal ($\mathrm{KTiOPO_4}$ or KTP) \cite{ViottiCLEO2019}. In this measurement, the d-scan setup is almost identical to the one shown in Fig.~\ref{fig:fewcyclesetup} with the only difference being the use of SF10 glass wedges, introducing approximately \SI{92}{fs^2/mm} of GVD at 1100\,nm (central wavelength of compressed pulses, dispersion window of \SI{920}{fs^2}), compared to only \SI{19}{fs^2/mm} for BK7 at that wavelength. The retrieved trace (Fig. \ref{fig:difscan}(e)) indicates \SI{21.4}{fs}-long pulses with a series of pre-pulses in the intensity profile, originating from uncompensated third-order dispersion.

For even longer, many-cycle pulses (>\SI{25}{fs}), the use of prism or grating compressors introduces appropriate amounts of dispersion (Yellow region in Fig. \ref{fig:vision}). Compressors which are integral parts of amplified, short pulse lasers can be conveniently used to perform d-scans. In Fig.~\ref{fig:difscan}(c) a d-scan measurement using a grating compressor as dispersive element is presented. The results were obtained with a Ti:Sapphire TW-class laser operated at 10\,Hz, driving a high intensity attosecond pulse beamline at LLC. 
One of the gratings in the compressor is mounted on a motorized translation stage that was continuously moved across the point of optimal compression. The dispersion of the compressor was evaluated to be \SI{4300}{fs^2/mm} of GVD. The total scanned dispersion window was \SI{17200}{fs^2} and the retrieval yields a pulse duration of \SI{43.4}{fs}.


 \section{Single-shot d-scan} 
 \label{sec:siscan}

We have discussed so far d-scan implementations where the dispersion variation was applied by mechanically moving an optical element inside the pulse compressor.  
For laser systems with high repetition rate (> 1 kHz) and pulse-to-pulse stability, this does not affect the accuracy of the pulse characterization. The obtained d-scan trace allows for retrieval of an average pulse in the pulse train.
However, for laser setups with low repetition rates or exhibiting shot-to-shot pulse duration fluctuations (which is rather common for TW-to-PW-level ultra high intensity systems), the solutions mentioned in the previous section can be either unpractical, take a long time to complete, or simply inaccurate in case of pulse instabilities.

Single-shot FROG implementations have emerged shortly after its introduction \cite{KaneOL1993}, while the architecture of SPIDER is fully compatible with single shot pulse measurements \cite{IaconisOL1998}. The first single-shot d-scan was demonstrated in 2015 \cite{FabrisOE2015}. 
In the following, the progress in the development of single-shot, SHG-based d-scan setups is reviewed and their performance in comparison to scanning d-scan approaches for few and multi-cycle light pulses is discussed.

\begin{figure}[hbt!]
     \centering
     \includegraphics[scale=0.8]{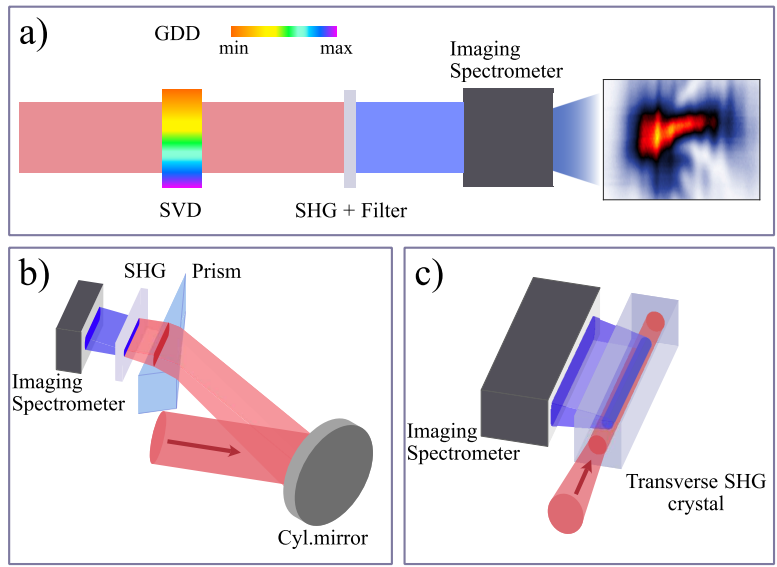}
     \caption{Principle of single shot d-scan measurement (a): SVD - spatially variable dispersion. Possible geometries for measuring few-(b) and multi-cycle (c) pulses.}
     \label{fig:sinshot}
 \end{figure}

In order to perform a single-shot measurement, all moving components should be eliminated from the optical setup. 
Two completely different approaches, presented in Fig.~\ref{fig:sinshot}, have been demonstrated so far: First, an optical element that encodes different amounts of GDD to different parts of the spatial beam profile was implemented. Second, a special nonlinear material that introduces both dispersion and nonlinearity was utilized. Common to both approaches is that the dispersion axis is translated into a spatial direction. 
The first approach is most conveniently implemented by replacing the scanning wedges in a standard d-scan with a prism that introduces spatially varying dispersion (SVD in Fig. \ref{fig:sinshot}(a)) over the beam profile.
After passing the prism, the light may be focused to a line into an SHG crystal, where now different positions along the line encode the SHG signal corresponding to different amounts of dispersion.
A d-scan trace is obtained, in single-shot, if the SHG signal along the line is imaged with an imaging spectrometer.

The first reported setup of that kind was designed to characterize few-cycle pulses from a hollow-core fiber compressor \cite{FabrisOE2015}. In this experiment, a slit was used to make a line which passed a BK7 prism. The output plane of the prism was imaged onto a thin BBO crystal, which is necessary to mitigate the angular dispersion which inevitably occurs at the backside of the prism. 
While this implementation is conceptually straightforward, the use of a slit for beam shaping can limit the SNR of the obtained d-scan traces and the setup is quite bulky.
The SNR can be improved by simply removing the slit and using the full beam profile, but a different focusing geometry is required. 
An elegant solution is to let the beam pass the prism first and then reflect off a spherical mirror under a large off-axis angle, introducing strong astigmatism. Adjusting the angles and distances between the prism, the mirror and the SHG crystal allows focusing the beam in one dimension while imaging the face of the prism in the other onto the crystal which results in a more compact and space-efficient design \cite{LouisyAO2017}.
A similar, but even simpler configuration, is depicted in Fig. \ref{fig:sinshot}(b), where the beam is focused with a cylindrical mirror to a line onto the SHG crystal, while the prism is placed in between.
The incidence angle on the mirror and the rotation of the prism have to be carefully aligned in order to minimize aberrations. Here, the angular chirp from the prism is not eliminated by imaging its output facet, but its impact is minimized by putting the SHG crystal directly after and as close to the prism as possible. As the beam is getting focused while propagating through the prism, care should be taken to avoid nonlinear effects in the prism. 
Finally, what is in common for all of the discussed implementations, is the need for a sufficiently homogeneous beam profile - significant intensity variations across the beam can decrease the accuracy of the measurement. In practice, a magnifying telescope and an iris can be used prior to the setup to select the central part of the beam profile for the measurement.

For the characterization of longer pulses, the method discussed above is no longer practical, as the amount of dispersion variation (e.g. the glass insertion window) that can be achieved for a reasonably large beam size in a single prism is limited to a few hundred \SI{}{fs^2} of GDD. 
An elegant alternative, that is also well-suited for longer pulses, is depicted in Fig. \ref{fig:sinshot}(c). In this implementation, a highly dispersive disordered nonlinear crystal (strontium barium nitrate, SBN) allowing for broadband transverse second harmonic generation (TSHG) is utilized both as the dispersive and nonlinear element \cite{Salgado-RemachaOL2020}.
The particular advantage of the randomly-ordered nonlinear crystal is its large dispersion, around \SI{500}{fs^2/mm}.
An initially negatively chirped pulse gradually compresses after entering the material and second harmonic is generated perpendicularly to the direction of propagation.
By recording the SHG with an imaging spectrometer, a d-scan trace is obtained in single-shot. 
For a typical crystal length of \SI{10}{mm}, a total dispersion window of \SI{5000}{fs^2} is obtained, allowing for measurements of many-cycle pulses with durations up to \SI{60}{fs} in the near-infrared spectral range \cite{Salgado-RemachaOL2020}.

To demonstrate the performance of single-shot d-scan implementations, we characterize near-single-cycle pulses after a HCF-based post-compression stage and multi-cycle pulses from standard mJ-level Ti:Sapphire CPA systems, using the geometries shown in Fig.~\ref{fig:sinshot}(b) and (c), respectively.
In both cases, the SHG signals were detected with home-made imaging spectrometers, using a compact, crossed Czerny-Turner design \cite{CzernyZfPh1930}, relying on divergent illumination of the grating in order to correct for astigmatism of the imaging path \cite{BatesJoPESI1970,AustinAO2009}.
More information on the spectrometer design, e.g. distances and angles between components, can be found in \cite{LouisyAO2017}. A cylindrical lens before the CCD sensor\cite{LeeOE2010} is added for additional aberration correction.

\begin{figure}[hbt!]
     \centering
     \includegraphics[width = \textwidth]{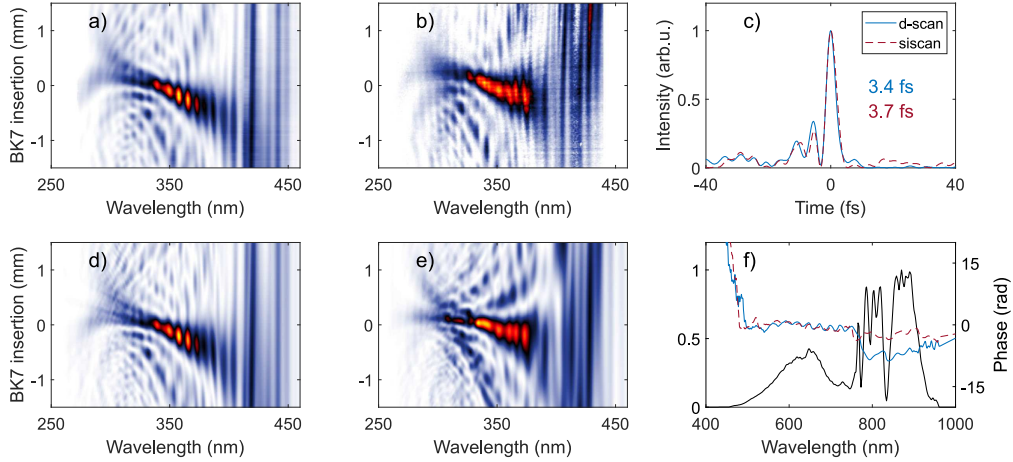}
     \caption{Measured traces using (a) standard and (b) single-shot setups for a HCF compressor system with retrievals shown in (d) and (e), respectively. (c) Retrieved pulse intensity profiles with FWHM durations indicated for both methods. (f) Measured spectrum and retrieved spectral phases. The blue lines are obtained with the scanning d-scan, while the red lines correspond to the single-shot measurements (siscan).}
     \label{fig:siscanresults}
 \end{figure}

 Figure \ref{fig:siscanresults} shows the results for the characterization of the few-cycle pulses.
 The d-scan traces obtained with the scanning (Fig. \ref{fig:siscanresults}(a)) and single-shot (Fig. \ref{fig:siscanresults}(b)) implementations are in good agreement, as confirmed by the retrieved pulse duration equal to \SI{3.4}{fs} and \SI{3.7}{fs} respectively.
  Both experiments show a slight tilt in the traces, indicating small amounts of uncompensated third-order dispersion, as also featured by the pre-pulses in the retrieved intensity profiles (Figure \ref{fig:siscanresults}(c)).
 The calculated RMS error $G$ was found to be equal to $1.5\ \%$ and $7\ \%$ for the scanning and single-shot measurements, respectively. In the frequency domain, the spectral phases agree quite well up to the wavelength $\lambda_s$ = \SI{766}{nm}, after which we observe  an almost constant relative shift of \SI{5.5}{rad}, which can be attributed to the low spectral amplitude at $\lambda_s$, introducing locally a high degree of uncertainty in the value of the phase.
 Essential for the agreement of the temporal profile however is not the phase itself, but its second-order derivative. A constant shift in the retrieved phase thus does not correspond to a different pulse.

\begin{figure}[hbt!]
     \centering
     \includegraphics[width = \textwidth]{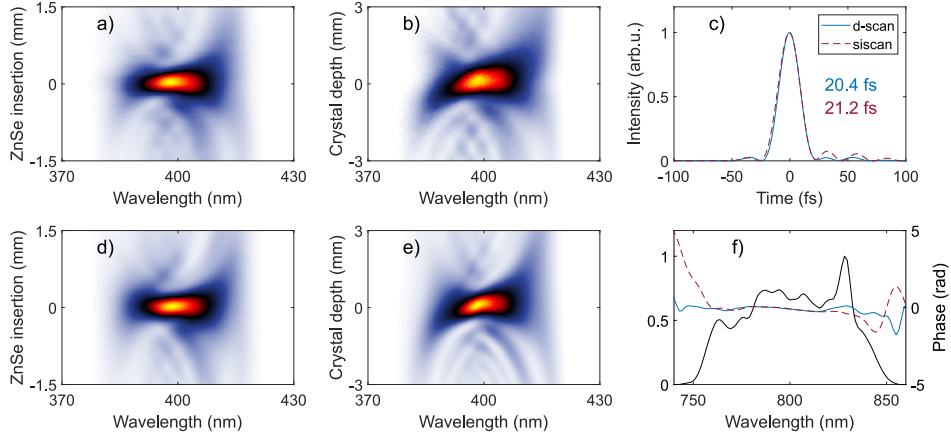}
     \caption{Measured traces (a) using standard and (b) single-shot setups for Ti:Sapphire CPA system with retrievals shown in (d) and (e), respectively. (c) Retrieved pulse intensity profiles with FWHM durations indicated for both methods. (f) Measured spectrum and retrieved spectral phases. The blue lines are obtained with the scanning d-scan, while the red lines correspond to the single-shot measurements (siscan).}
     \label{fig:siscanresultsatto}
 \end{figure}

The results from the implementation based on the random nonlinear crystal (Fig.~\ref{fig:sinshot}(c)) are summarized in Fig.~\ref{fig:siscanresultsatto}.
The measurements were performed with a 1\,kHz, mJ-level, Ti:Sapphire CPA laser system at LLC, emitting near transform-limited pulses with duration around 20 fs (FHWM).
The conventional (scanning) d-scan measurement, shown in Figure \ref{fig:siscanresultsatto}(a), utilized a pair of ZnSe wedges, featuring extremely large dispersion (GVD = \SI{1025}{fs^2/mm}) in the near-infrared.
An acousto-optic programmable dispersive filter (Dazzler, Fastlite), which is an integral part of the CPA chain, introduced negative chirp.
The single-shot setup used a 10-mm long SBN crystal with group velocity dispersion of \SI{480}{fs^2/mm} at 800\,nm. The results are shown in panel (b). Again, the experimental traces are in good agreement, while the difference in width can be attributed to slightly different range of dispersion windows and a larger amount of residual TOD for the single-shot measurement.
The pulse retrieval results agree very well between the two setups  in terms of the temporal intensity profiles (Fig.\ref{fig:siscanresultsatto}(c)) and retrieved spectral phases (Fig. \ref{fig:siscanresultsatto}(d)) (RMS error $0.4\ \%$ for scanning d-scan and $1.9\ \%$ for the single shot version).

\section{Conclusion}
In this work, we presented a review of the progress of pulse characterization using the second harmonic dispersion scan technique.
We show that the traces obtained with d-scan are naturally very intuitive to interpret, with different polynomial contributions to the pulse spectral phase appearing as characteristic deformations of the trace. We also give a brief orientation in the available phase reconstruction algorithms which can be implemented to retrieve the exact pulse information.
By employing different pulse compressor configurations, d-scan can be successfully adapted to the measurement of pulses with different pulse durations and central frequencies.
Furthermore, we present two different single-shot implementations, well-suited for the characterization of pulse sources with low repetition rate or substantial pulse-to-pulse fluctuations, where conventional (scanning) d-scan would either take inconveniently long time or result in a misleading conclusion.

While the d-scan so far has been primarily used for the characterization of pulses in the near infrared spectral range, derived from Ti:Sapphire or Ytterbium-doped lasers, the adaption to other wavelength ranges is straight-forward.  
In recent years, there has been a great deal of progress in the development of light sources providing ultrashort pulses in the short, mid, and long-wave infrared spectral regions as well as the deep UV \cite{LiNC2017,PupeikisOptica2020,AndriukaitisOL2011,BockJOSAB2018,ReiterOL2010,GalliOL2019}.
Expanding the d-scan technique to different carrier wavelengths is a subject of ongoing research (see e.g. \cite{TajalliOL2019} for the UV range) and without doubt we will see more work in this direction in the future.

\section*{Funding}

The authors acknowledge support from the Swedish Research Council (2013-8185, 2016-04907, 2019-06275); the European Research Council (proof of concept grant SISCAN-789992); the Knut and Alice Wallenberg Foundation; Junta de Castilla y León (SA287P18); Ministerio de Economía y Competitividad (EQC2018-004117-P, FIS2017-87970-R); PT2020 (program 04/SI/2019 Projetos I\&D industrial \'{a} escala grant no. 045932); Funda\c{c}\~{a}o para a Ci\^{e}ncia e a Technologia (FCT)('Ultragraf' M-ERA- NET4/0004/2016).  

\section*{Disclosures}
B. A.: USAL (P), C. A.: SPH (I,P), H. C.: SPH (I,C,P,R), P. T. G.: SPH (E, P, R), C. G.: LU (P), A. L.: SPH (I, P), M. M.: SPH (I, E, P, R), R. R.: SPH (I, E, P, R). LU, Lund University, SPH, Sphere Ultrafast Photonics, USAL,University of Salamanca
\bibliography{Ref_lib_test}
\end{document}